# Cryogenic rf test of the first plasma etched SRF cavity

J. Upadhyay,[a] A. Palczewski,[b] S. Popović,[a] A.-M. Valente-Feliciano,[b] D. Im,[a] L. Phillips,[b] and L. Vušković[a]

[a]Department of Physics and Center for Accelerator Science, Old Dominion University, Norfolk, VA 23529, USA

[b]Thomas Jefferson National Accelerator Facility, Newport News, VA 23606, USA

**Abstract**

Plasma etching has a potential to be an alternative processing technology for superconducting radio frequency (SRF) cavities. An apparatus and a method are developed for plasma etching of the inner surfaces of SRF cavities. To test the effect of the plasma etching on the cavity rf performance, a 1497 MHz single cell SRF cavity is used. The single cell cavity is mechanically polished, buffer chemically etched afterwards and rf tested at cryogenic temperatures for a baseline test. This cavity is then plasma processed. The processing was accomplished by moving axially the inner electrode and the gas flow inlet in a step-wise manner to establish segmented plasma processing. The cavity is rf tested afterwards at cryogenic temperatures. The rf test and surface condition results are presented.

## 1. Introduction

Superconducting radio frequency (SRF) cavities are integral components of accelerators used in nuclear and high energy physics research. Currently, the inner surfaces of SRF cavities are chemically treated (etched or electro-polished) to remove impurities, mechanically damaged layers and reduce the surface resistance of the superconducting surface, thus achieve a favorable rf performance. These technologies are based on the use of hydrogen fluoride (HF) in liquid acid baths [1-6], which poses a major environmental and personal safety concern. The plasma etching method would present a much more controllable, less expensive, and more environment-friendly processing technology. This competitive alternative would also provide the unique opportunity to modify the niobium (Nb) surface for energy efficient superconducting rf properties.

The plasma etching method described here uses a coaxial capacitive radiofrequency discharge of $Ar/Cl_2$ mixture operating at 13.56 MHz. The $Cl_2$ gas used in the process forms volatile compounds in reaction with the Nb and its oxides in an rf plasma environment. Before plasma etching a single cell SRF cavity, ring type Nb samples were used in a single-diameter and a varied-diameter cylindrical cavity to measure the effect of the process parameters on the Nb etching as reported in Refs. [7, 8]. Applying a positive dc bias to the inner electrode and changing the contour of the inner electrode compensated the sheath voltage asymmetry due to a lower surface area of the inner electrode in the coaxial plasma. The etch rate dependence on pressure, rf power, $Cl_2$ concentration and diameter of the inner electrode was measured and reported in Ref. [7]. The etch rate dependence on the temperature, dc bias and understanding of the etch mechanism was reported in Ref. [8]. It was found that there is a strong etch rate non-uniformity in the direction of the gas flow. Its dependence on the process parameters is also reported in Ref. [8]. The concept of surface enhancement of the inner electrode to partially overcome sheath voltage asymmetry is applied and various structures were tested [9]. The optimum, corrugated structure for reversal of the asymmetry has been determined [9]. A stainless steel pillbox cavity was chosen with the aim to study the plasma processing effect on varied-diameter structures, as uniform plasma-surface interaction is a challenging task [10]. The segmented plasma production by moving the inner electrode and gas flow inlet in a stepwise manner is chosen. The apparatus and method developed for segmented plasma production is reported in Ref. [11]. The cryogenic rf test of an actual plasma etched SRF cavity is the only thing left to compare to plasma etching with chemical etching technologies. The approach to compare rf performance was to use a single cell SRF cavity, perform the buffer chemical processing (BCP) and apply the cryogenic rf test. The same cavity was then plasma etched and retested at cryogenic temperature.

## 2. Experiment and method

The apparatus, shown in Fig. 1, was used to plasma etch a single cell SRF cavity. It consisted of an rf power supply, which was equipped with a matching network and connected in series to a dc power supply that provided a positive dc bias to the inner electrode. RF power is coupled to the inner (driven) electrode with a coaxial atmospheric pressure feedthrough. The feedthrough and the inner electrode are attached to a controllable axially moving manipulator, which is shown on the left side of Fig. 1. The cavity, acting as a vacuum vessel, is connected through the antechamber to the pumping system, which consists of a turbo molecular vacuum pump and a roughing pump with vacuum valves and diagnostic gauges. The exhaust gases are collected and processed in a homemade scrubber that is filled with sodium hydroxide solution in water. Gas is fed to the system through a mixing manifold and a specially designed gas inlet, which disperses the gas mixture. The gas inlet is connected to a second controllable axially moving manipulator, which is synchronized with the first manipulator, which is shown in the right side of the image. The gas inlet is a double conical shaped structure as described in Ref. [11]. The gas flow inlet was a part of the experimental setup, and it was electrically grounded. The inner electrode was corrugated and made of stainless steel. It was 9.0 cm long, which is smaller than the length of the cell structure.

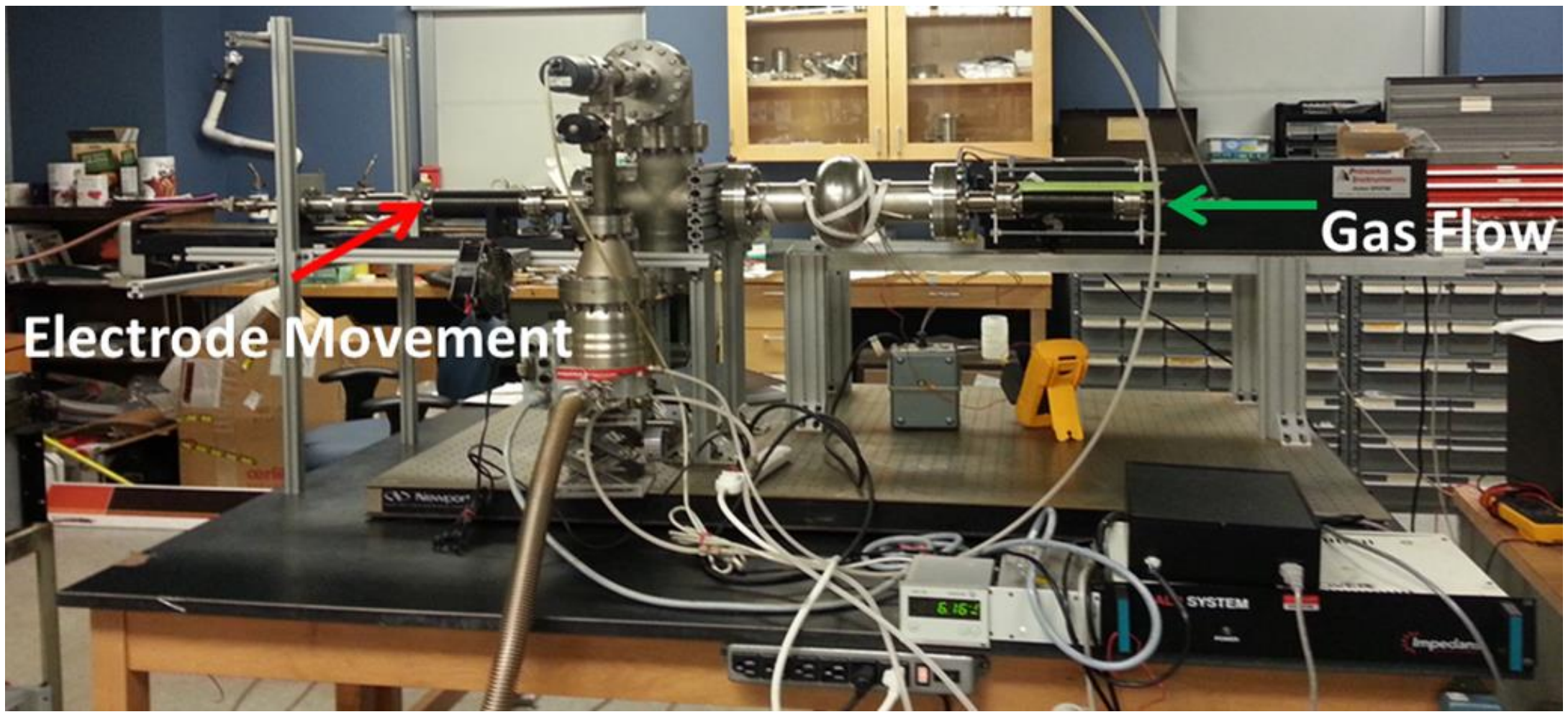

FIG. 1. The experimental setup to plasma etch single cell SRF cavity.

The cavity wall was electrically grounded and served as the outer electrode of the cylindrical rf discharge. A heating tape was wrapped around the cavity and it was attached to a variable-voltage transformer to control the surface temperature during the etching process. The surface temperature was monitored with the help of a thermocouple and a multi-meter.

## I. RESULTS AND DISCUSSIONS

A single cell 1497 MHz SRF cavity was mechanically polished and then buffered chemically etched for additional 60 micron material removal. The single cell cavity was then heat treated in a vacuum furnace at $600^0$ C for 10 hours, degreased and high-pressure water rinsed. The cavity was tested at cryogenic temperatures at Jefferson Lab VTA facility. The $Q_0$ vs. $E_{acc}$ test results are shown in the Fig. 2. The details about test methods can be founded in Ref. [12].This BCP cavity shows field emission around 15 MV/m as reported by the curve with the date (8/27/2014).

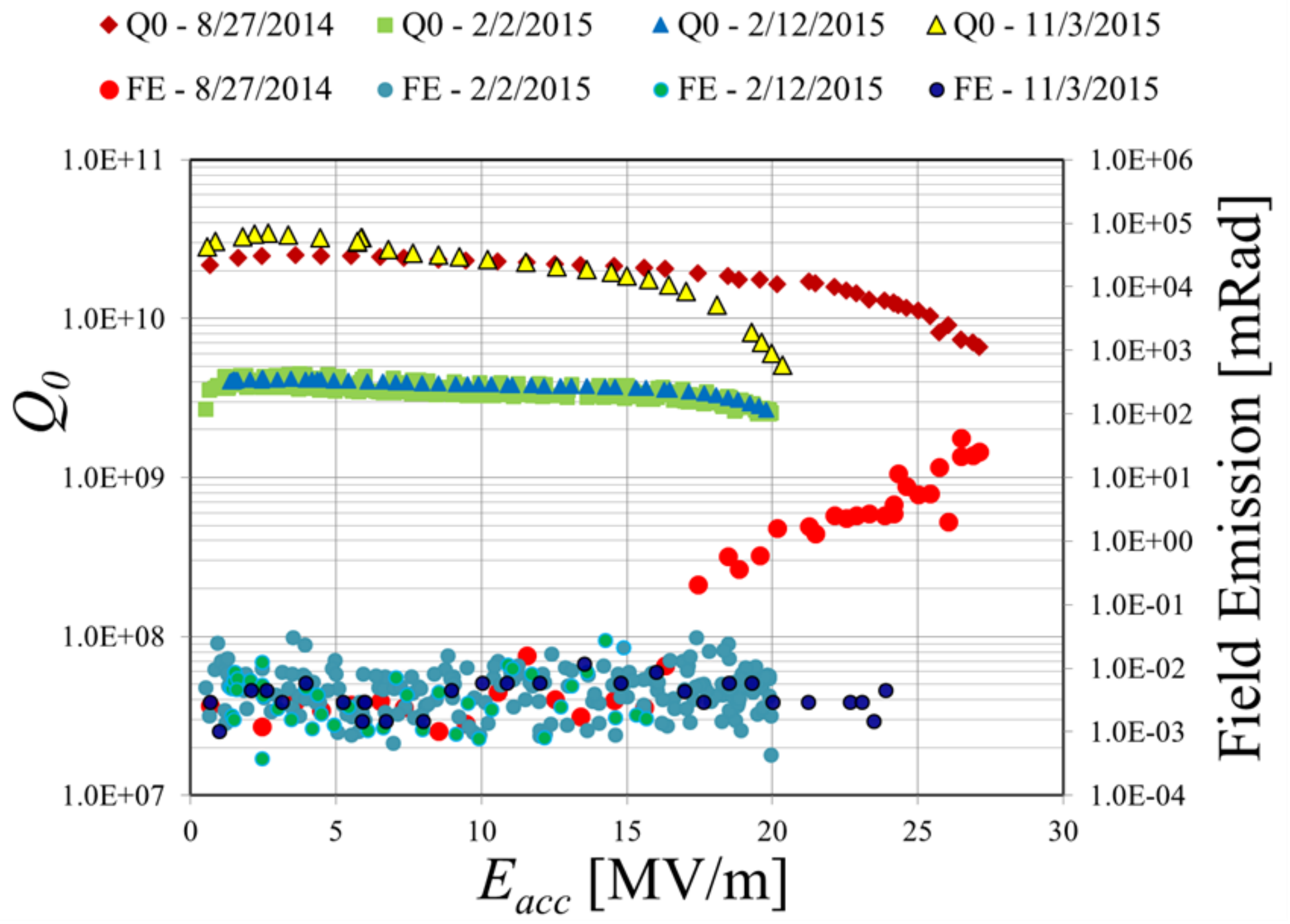

FIG. 2. The rf test results of the plasma etched SRF cavity at 1.8 K. Top portion of the graph presents $Q_0$ and the bottom portion presents field emission.

The cavity was then plasma etched for 24 hours at Old Dominion University. The conditions during plasma etching were the following: pressure 50 mTorr, rf power 160 W, dc bias 320 V, temperature 231 $^0$C and dc current 0.930 A. The gas flow rate was 0.43 l/min and the gas mixture used was 15% $Cl_2$ mixed with Argon. The uncertainty of $Cl_2$ concentration was 2%, in rf power 10 W, in pressure 4 mTorr, in dc bias 10 V and in dc current 10 mA. The Nb removed from the cell structure was in the order of 10 µm, while the material removed from the beam tube was in the order of 100 µm. The separation between inner electrode and gas flow inlet during plasma etching process was kept constant at 5 cm. The powered electrode was positioned at the beginning of the etching at one end of the cell and moved sequentially towards the other end of the cell. The beam tube plasma etching occurred due to the expansion of the plasma produced during cell etching.

After 24 hours of plasma etching, the cavity was kept at temperature for 10 additional hours and the vacuum pumping system was active for an additional 8 hours. The cavity was then open to atmospheric pressure. A thick black residue was found on the cavity surface. This residue was collected and analysed. The post plasma etched cavity looked heavily contaminated as shown in Fig. 3.

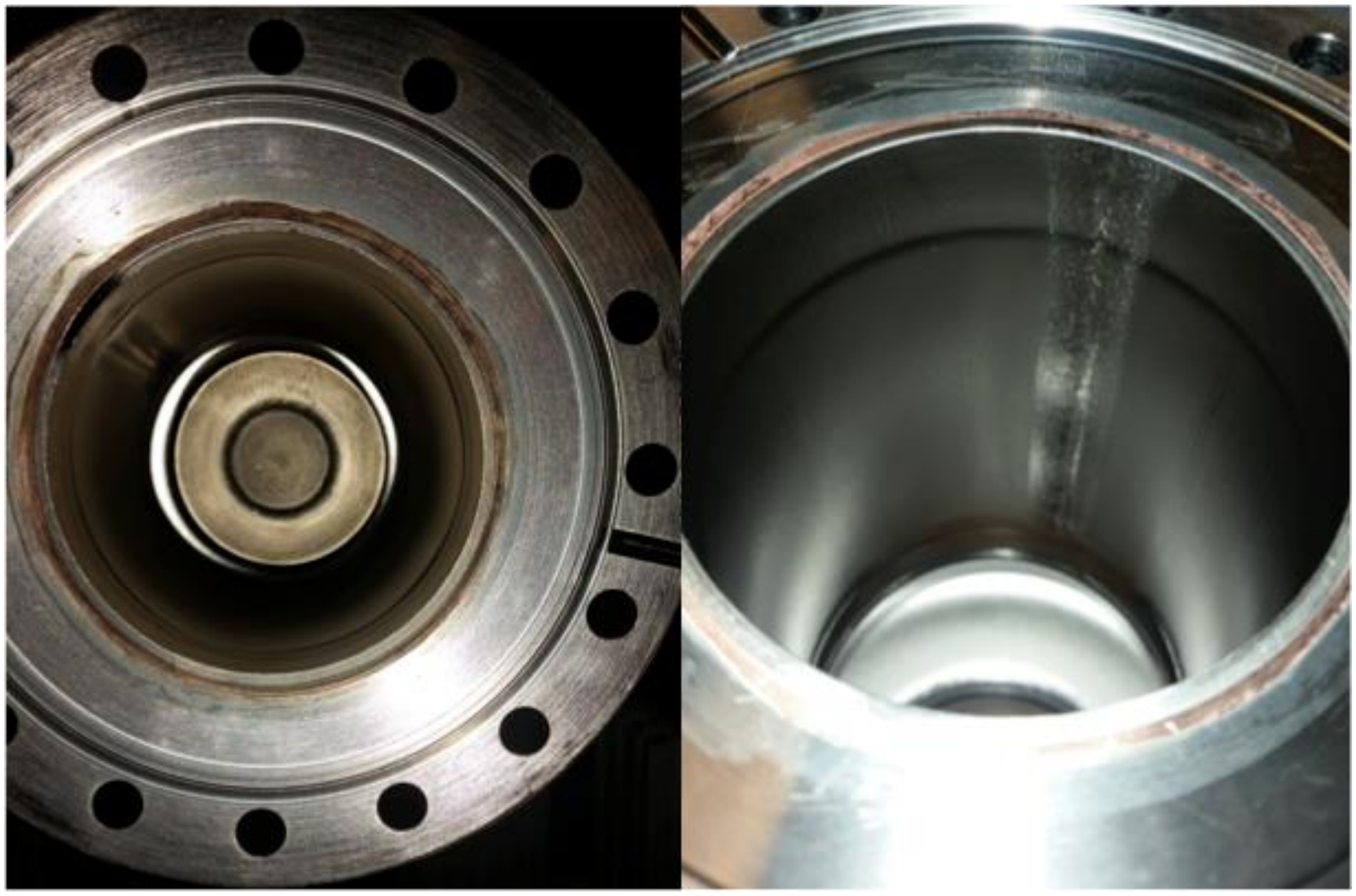

FIG. 3. The images of single cell cavity plasma etched (left) and buffer chemical polished (right).

Surface analysis of the residue shows it to be Fe, Ni, Cr and $Cl_2$, which suggest that the inner electrode and gas flow inlet, which were made of stainless steel were plasma etched and fell on the SRF cavity surface. The visual inspection of gas flow inlet and the elemental analysis of the residue confirm this. The images from the surface analysis are shown in Fig. 4.

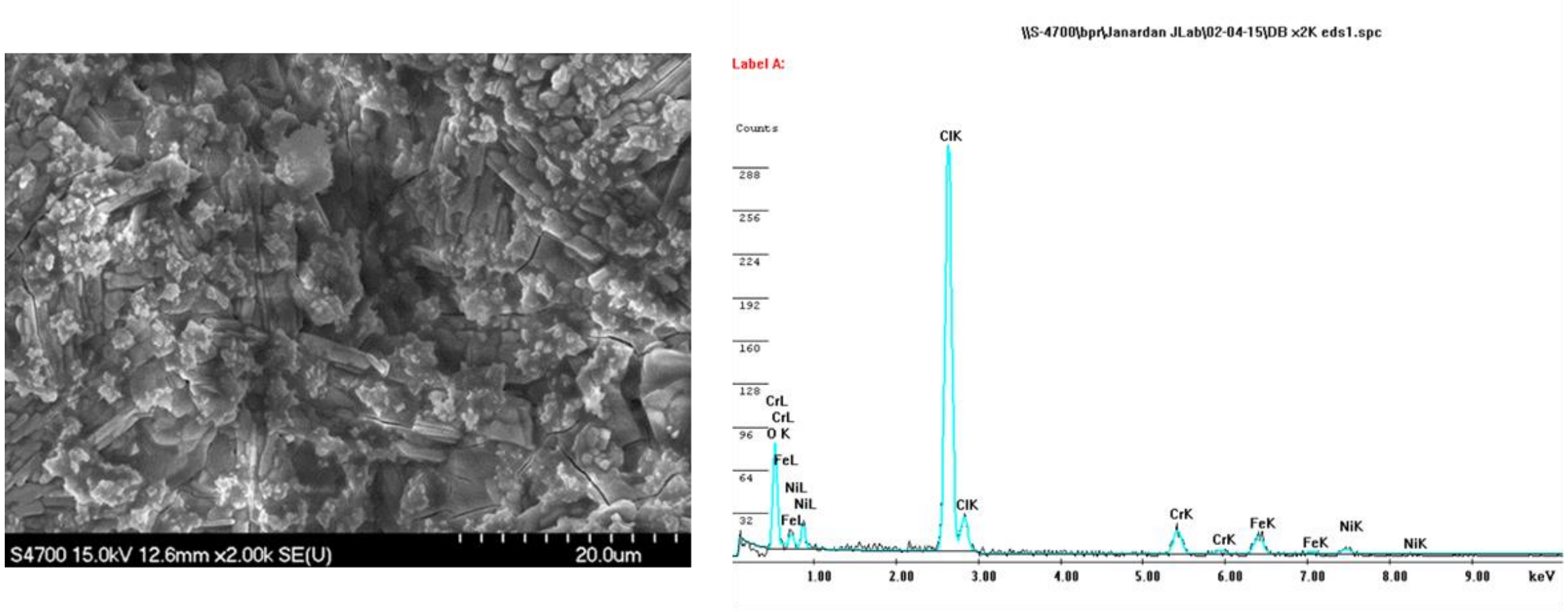

FIG. 4. The images of surface analysis (left) and elemental composition (right) of the residue from plasma etched cavity.

The plasma etched cavity was water rinsed, ultrasonically cleaned and high pressure water rinsed at Jefferson lab. The cavity was then rf tested at cryogenic temperatures. The test result has shown the quality factor is reduced by an order of magnitude compared to BCP cavity and quenching at around 20 MV/m though field emission shows significant reduction as shown by the curve (2/2/2015) in Fig. 2. There were two strong possibilities for this degradation in the quality factor; one could be the deposition of non-superconducting material on the cavity surface, the other would be a large amount of hydrogen present in the bulk, which decreases the quality factor and known as a Q disease.

To test for Q-disease, the cavity was kept at 90K to 140 K over 14 hours and tested. The test result shows no hydrogen disease, as the Q curve looks exactly the same as fast cooled down cavity Q curve as shown by the curve in Fig.4 by (2/12/2015).

To test the possibility of the stainless steel residue being the cause of the Q degradation in the cavity it was chemically cleaned with an aqua regia solution, and phosphoric acid. The use of HF was avoided in order not to disturb the Nb oxide surface and protect the Nb from any etching. During the phosphoric acid cleaning the temperature of the acid was raised to $100^0$ C for 60 minute. The cavity was degreased and high-pressure water rinsed and tested again. The removal of stainless steel residue helped and the quality factor of the cavity came back to the BCP cavity level. There was no sign of field emission during this chemically cleaned plasma etched cavity test as shown in Fig. 4 (11/3/2015). A similar observation on reduction of field emission due to plasma surface interaction was observed and reported for flat samples in Ref. [13] and for the SRF cavity in Ref. [14].

Therefore, the earlier degradation of Q factor was due to stainless steel residue deposition on the surface of the SRF cavity. The conclusion is that all components (inner electrode, gas flow inlet) used in future plasma etching apparatus should be made of Nb and electrically biased to prevent these components from plasma etching. All the etched Nb was not removed from the system due to huge amount of material etched and partial condensation on some surfaces. Raising the temperature and using high pumping speed could improve the purity of the cavity surface.

## CONCLUSION

We presented the experimental setup and procedure to etch a single cell SRF cavity in an rf plasma discharge and the rf test results of the first plasma etched SRF cavity at cryogenic temperature. The test results suggest that the plasma-etched cavity would perform as good as chemically etched cavity if the component used during the processing are made of Nb or electrically isolated, so that the processing plasma should not etch the components. The plasma etched cavity has shown no field emission. Field emission did not increase even after multiple chemical cleaning and testing. This very first test result of a plasma etched cavity shows a viable, environment friendly and less expensive technology compared to wet etching technology. It also holds the promise to implement the tailoring of the surface for better superconducting properties.

## I. ACKNOWLEDGMENT

This work is supported by the Office of High Energy Physics, Office of Science, Department of Energy under Grant No. DE-SC0014397. Thomas Jefferson National

Accelerator Facility, Accelerator Division supports J. Upadhyay through fellowship under JSA/DOE Contract No. DE-AC05-06OR23177.

**References**

[1] P. Kneisel, *Nucl. Instr. Meth. Phys. Res. A* **557**, 250 (2006).

[2] K. Saito, H. Inoue, E. Kako, T. Fujino, S. Noguchi, M. Ono, and T. Shishido, Proceedings of the 1997 workshop on RF superconductivity, Abano Terme (Padova), Italy pp 795.

[3] L Lilje, C Antoine, C. Benvenuti, D. Bloess, J.-P. Charrier, E. Chiaveri, L. Ferreira, R. Losito, A. Matheisen, H. Preis, D. Proch, D. Reschke, H. Safa, P. Schmuser, D. Trines, B. Visentin, and H. Wenninger, *Nucl. Instr. Meth. Phys. Res. A* **516**, 213 (2004).

[4] H. Tian and C. E. Reece, *Phys. Rev. ST Accel. Beams* **13**, 083502 (2010).

[5] C. A. Cooper, and L. D. Cooley, *Supercond. Sci. Technol.* **26**, 015011 (2013).

[6] A. D. Palczewski, C. A. Cooper, B. Bullock, S. Joshi, A. A. Rossi, and A. Navitski, Proceedings of SRF 2013, Paris, France.

[7] J. Upadhyay, D. Im, S. Popović, A.-M. Valente-Feliciano, L. Phillips, and L Vusković, *Phys. Rev. ST Accel. Beams* **17**, 122001 (2014).

[8] J. Upadhyay, D. Im, S. Popović, A.-M. Valente-Feliciano, L. Phillips, and L Vusković, *J. Appl. Phys.* **117**, 113301 (2015).

[9] J. Upadhyay, D. Im, S. Popović, A.-M. Valente-Feliciano, L. Phillips, and L Vusković, *J. Vac. Sci. Technol. A* **33 (2)**, 061309 (2015).

[10] A. Hershcovitch, M. Blaskiewicz, J. M. Brennan, A. Custer, A. Dingus, M. Erickson, W. Fischer, N. Jamshidi, R. Laping, C.-J. Liaw, W. Meng, H. J. Poole, and R. Todd, *Phys. Plasmas* **22**, 057101 (2015).

[11] J. Upadhyay, D. Im, J. Peshl, M. Basovic, S. Popović, A.-M. Valente-Feliciano, L. Phillips, and L. Vusković, *Nucl. Instr. Meth. Phys. Res. A* **818,** 76 (2016).

[12] T. Powers, Proceedings of the 12th International Workshop on RF Superconductivity, Cornell University, Ithaca, New York, USA.

[13] P.V. Tyagi, M. Doleans, B. Hannah, R. Afanador, C. McMahan, S. Stewart, J. Mammosser, M. Howell, J. Saunders, B. Degraff, S-H. Kim, *Applied Surface Science* **369**, 29 (2016).

[14] M. Doleans, , P.V. Tyagi, R. Afanador, C.J. McMahan, J.A. Ball, D.L. Barnhart, W. Blokland, M.T. Crofford, B.D. Degraff, S.W. Gold, B.S. Hannah, M.P. Howell, S-H. Kim, S-W. Lee, J. Mammosser, T.S. Neustadt, J.W. Saunders, S. Stewart, W.H. Strong, D.J. Vandygriff, D.M. Vandygriff, *Nucl. Instr. Meth. Phys. Res. A* **812,** 50 (2016).